\documentclass[12pt,a4paper]{article}
\pdfoutput=1
\usepackage{jcappub}
\usepackage{bm}
\usepackage{bbm}
\usepackage{etoolbox}
\usepackage{mhchem}
\usepackage[english]{babel}
\usepackage{multirow}
\usepackage{url}
\usepackage{caption}
\usepackage{subcaption}
\usepackage[utf8]{inputenc}
\usepackage[dvipsnames]{xcolor}

\usepackage{graphicx}
\usepackage{xcolor}
\usepackage{amsmath,amssymb}
\usepackage{bm}
\usepackage{hyperref}
\usepackage{url}
\usepackage{color}
\usepackage{fleqn}
\usepackage{dcolumn}

\def\bea{\begin{eqnarray}}
\def\eea{\end{eqnarray}}
\def\be{\begin{equation}}
\def\ee{\end{equation}}
\newcommand{\Sec}[1]{Sec.~\ref{#1}}

\makeatletter
\gdef\@fpheader{}
\makeatother

\title{Robust bounds on ALP dark matter from dwarf spheroidal galaxies \\
in the optical MUSE-Faint survey}

\author[a,b]{Elisa Todarello,} 
\emailAdd{elisamaria.todarello@unito.it}
\author[a,b]{Marco Regis,} 
\author[a,b]{Javier Reynoso-Cordova,}%
\author[b]{Marco Taoso,}
\author[c,d]{Daniel Vaz,}
\author[c,e]{Jarle Brinchmann,}
\author[f]{Matthias Steinmetz,}
\author[e]{Sebastiaan L. Zoutendijk}

\affiliation[a]{Dipartimento di Fisica, Universit\`{a} di Torino, via P. Giuria 1, I--10125 Torino, Italy}
\affiliation[b]{Istituto Nazionale di Fisica Nucleare, Sezione di Torino, via P. Giuria 1, I--10125 Torino, Italy}
\affiliation[c]{Instituto de Astrofísica e Ciências do Espaço, Universidade do Porto, CAUP, Rua das Estrelas, PT4150-762 Porto, Portugal}
\affiliation[d]{Departamento de F{\'\i}sica e Astronomia, Faculdade de Ci{\^e}ncias, Universidade do Porto,
Rua do Campo Alegre 687, PT4169-007 Porto, Portugal}
\affiliation[e]{Leiden Observatory, Leiden University, P.O.~Box~9513, 2300~RA~Leiden, The Netherlands}
\affiliation[f]{Leibniz-Institut f\"ur Astrophysik Potsdam (AIP), An der Sternwarte 16, 14482 Potsdam, Germany}

\abstract{
Nearby dwarf spheroidal galaxies are ideal targets in the search for indirect dark matter (DM) signals.
In this work, we analyze MUSE spectroscopic observations of a sample of five galaxies, composed of both classical and ultra-faint dwarf spheroidals. The goal is to search for radiative decays of axion-like particles (ALPs) in the mass range of 2.7-5.3 eV. After taking into account the uncertainties associated with the DM spatial distribution in the galaxies, we derive robust bounds on the effective ALP-two-photon coupling. 
They lie well below the QCD axion band and are significantly more constraining than limits from other probes, in the relevant mass range.
We also test the possible presence of a positive signal, concluding that none of the wavelength channels selected for this analysis, i.e., not affected by large background contamination, is exhibiting such evidence. 
}

\date{\today}

\begin{document}
\maketitle

\section{\label{sec:intro} Introduction}

Axion-like particles (ALPs) are pseudo Nambu-Goldstone bosons that arise in extensions of the Standard Model and that can act as cold dark matter (DM) candidates~\cite{Preskill:1982cy,Abbott:1982af,Dine:1982ah}.
A particularly well-motivated example is the QCD axion, which is associated with the Peccei-Quinn symmetry solution to the strong CP problem~\cite{Peccei:1977hh,Peccei:1977ur, Weinberg:1977ma, Wilczek:1977pj}. 
Generically, ALPs couple to photons through the operator $\mathcal{L}=-\frac{1}{4}g_{a\gamma}\,a\,F_{\mu\nu}\tilde{F}_{\mu\nu},$ where $a$ is the ALP field, $F_{\mu\nu}$ is the electromagnetic field strength, $\tilde{F}_{\mu\nu}$ its dual, and $g_{a\gamma}$ the coupling constant. This interaction term leads to a variety of possibilities to detect ALPs in laboratory experiments or with astrophysical and cosmological probes, see e.g.~\cite{Graham:2015ouw,Irastorza:2018dyq,DiLuzio:2020wdo} for reviews. 
In astrophysical environments, an almost monochromatic photon emission is produced by the radiative decay of ALP DM, and,
for ALP masses in the eV range, this photon line falls in the optical and near-infrared bands.
In this frequency range, several upper bounds on this signal have been derived from observations~\cite{Ressell:1991zv,Bershady:1990sw,Grin:2006aw,Regis:2021,Nakayama:2022jza,Carenza:2023qxh, Bessho:2022yyu, Yin:2023uwf}. In particular, ref.~\cite{Regis:2021} derived the currently most stringent constraints on ALP radiative decays for masses between 2.7 and 5.3 eV, improving previous bounds by more than an order of magnitude. Interestingly, in recent years, ALPs masses in the eV mass range have been invoked to explain excesses in the measured cosmic near-infrared background and its angular anisotropies~\cite{Gong:2015hke,Kalashev:2018bra,Caputo:2020msf,Bernal:2022wsu,Nakayama:2022jza,Carenza:2023qxh}. 
The upper limits of~\cite{Regis:2021} severely challenge some of these scenarios~\cite{Caputo:2020msf}.

The analysis in~\cite{Regis:2021} is based on spectroscopic observations of the Leo T dwarf spheroidal galaxy obtained with the Multi Unit Spectroscopic Explorer (MUSE) at the Very Large Telescope (VLT)~\cite{MUSE2017}.
Dwarf spheroidal galaxies are ideal targets for searching for DM decay signals because they contain large DM densities and are relatively close to us. Moreover, measurements of the line-of-sight velocity of the stars in these objects allow us to infer the underlying DM distribution along with its uncertainty. This information is instrumental in order to reliably predicting the ALP signal, and deriving robust bounds on the ALP decay lifetime.

With the present work, we extend and improve the analysis in \cite{Regis:2021} in two ways: we enlarge the dataset exploiting recent MUSE observations of other dwarf galaxies, and we implement a more detailed treatment of the DM distribution and its uncertainty,
taking advantage of recent analyses. 
More specifically, in addition to Leo T, we consider MUSE observations of the dwarf spheroidal galaxies Sculptor, Eridanus 2, Grus 1, and Hydra II.
Then, for Eridanus 2, Grus 1, Hydra II, and Leo T, we make use of the 
recent determination of the DM content in these objects performed by the MUSE collaboration~\citep{
Bas:2021}.
Concretely, we consider two parametrizations for the DM density, namely the Navarro–Frenk–White (NFW) model~\cite{Navarro:1995iw} and a cored profile.
We account for the uncertainty on these DM distributions including the corresponding likelihoods derived in~\citep{Bas:2021} in our statistical analysis; see \Sec{sec:res} for details.
For Sculptor, we follow the same procedure but we derive the DM distribution and the relevant likelihood ourselves. This is accomplished by means of a Jeans analysis, using data from~\cite{2009AJ....137.3100W,Coleman_2005} and employing the same method as~\citep{Bas:2021}.

For each target, we perform a search of ALP decay signals in the MUSE data, and then we combine the individual bounds in a global analysis. 
We find upper limits on the ALP lifetime similar to, but slightly weaker than, those in~\cite{Regis:2021}. 
Finally, excluding channels severely contaminated by background, we do not find significant evidence for an ALP signal.

The structure of this paper is as follows. The data from MUSE observations are presented in~\Sec{sec:data}. The calculation of the ALP decay signal is discussed in~\Sec{sec:axion}. In~\Sec{sec:res}, we discuss the statistical analysis and results. We conclude in~\Sec{sec:conc}. The Jeans analysis for Sculptor is discussed in Appendix~\ref{sec:scu}. The caveats of our analysis assuming a cored dark matter profile are addressed in Appendix~\ref{sec:cav}.

\section{\label{sec:data} Observations and data reduction}
As part of MUSE-Faint, a GTO survey of faint dwarf galaxies (PI Brinchmann), Leo T, Sculptor, Eridanus 2, Grus 1, and Hydra II\footnote{Sculptor,  which contains a significantly larger number of stars than the other targets, is commonly considered to be a ``classical" dwarf, while Leo T, Eridanus 2, Grus 1, and Hydra II are classified as ``ultra-faint" dwarf galaxies.} were observed with MUSE, a large-field medium-resolution Integral Field Spectrograph installed on the VLT. We use multiple exposures of 900~s of each galaxy, with a total exposure time of 3.75 hours on Leo T (one field), 3 hours on Sculptor (one field), 21.5 hours on Eridanus 2 (five fields), 4 hours on Grus 1 (one field), and 14.75 hours on Hydra II (four fields).

The data were taken in the Wide Field Mode with adaptive optics (WFM-AO), which provides a $1 \times 1~\mathrm{arcmin^2}$ field of view with a spatial sampling of $0.2~\mathrm{arcsec~pixel^{-1}}$. 
The data cover a wavelength range of $4700 -  9350~\rm{\AA}$, sampled at a resolution of $1.25~\rm{\AA}$.  A blocking filter was used to remove the light from the sodium laser of the adaptive optics system to avoid contamination. This filter blocked light in the $5820-5970~\rm{\AA}$ ($2.13 - 2.08$ eV) range, which appears as a gap in the constraints presented in the following.

For data reduction, we refer the reader to Ref.~\cite{Bas:2021} for details, while here we provide a brief summary. We performed the standard data reduction procedure using
the MUSE Data Reduction Software (DRS; version 2.8~\cite{2020arXiv200608638W}). Flux calibration was carried out using flux standards observed during the night, while atmospheric emission lines were removed by accounting for Raman scattering caused by the laser light of the adaptive optics system.  We subtracted emission lines from the night sky that have well-known wavelengths and result in increased noise at those wavelengths. We measured a spatial resolution (full-width half maximum) of 0.61, 0.50, 0.53, 0.67, and 0.40 arcsec for Leo T, Sculptor, Eridanus 2, Grus 1, and Hydra II, respectively, at a wavelength of $7000~\rm{\AA}$ in the reduced datacubes.

To ensure an accurate analysis of the data cubes, it is crucial to have a reliable estimate of the noise. Previous studies (e.g., Ref~\cite{2017A&A608A1B}) have shown that the MUSE Data Reduction Software (DRS) underestimates uncertainties in the final data cube. Therefore, we proceeded as in Ref.~\cite{Regis:2021}  and re-estimated the pixel-to-pixel variance directly from each individual exposure data cube using the method described in Ref.~\cite{2017A&A608A1B}, creating mask images using SExtractor \cite{1996A&AS..117..393B}.
We then combined all single-exposure data cubes using MPDAF \cite{2017arXiv171003554P}, to create two final data cubes that were used in the subsequent analysis,  one for the r.m.s. noise $\sigma_{rms}$ and one for the observed flux density $S_{obs}$.

\begin{figure}[t!]
\centering
  \includegraphics[width=\textwidth]{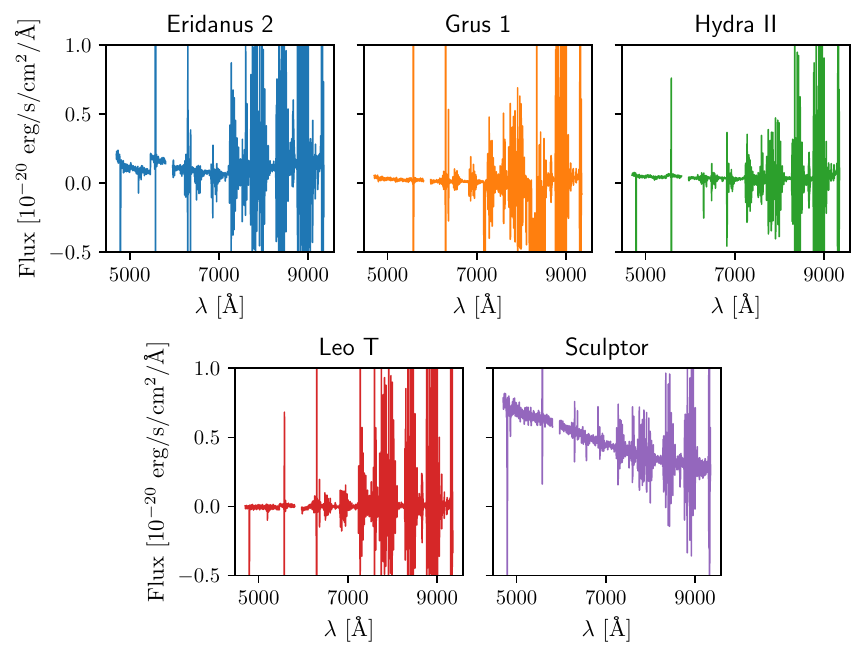}
    \caption{Flux density per beam solid angle averaged over the unmasked pixels of the map as a function of the wavelength of observation.}
\label{fig:illustr}
 \end{figure}

The data contain numerous stellar sources within the field of view, both from the dwarf galaxy and also from some likely foreground stars from the Milky Way, as well as some galaxies. 
To minimize the impact of these sources on the final results, we have identified and masked the brightest ones.
This was achieved by following the same approach as in Ref.~\cite{Regis:2021} and involved two steps. First, we generated a white-light image by summing over the wavelength axis in each datacube. Next, we ran SExtractor on this white-light image, with a detection threshold of $3\sigma$, resulting in a segmentation map that was used to mask sources.  Therefore, we consider only pixels where no sources are detected in the white-light image.
In Fig.~\ref{fig:illustr} we show, for all the dwarf galaxies under consideration, the flux density per beam solid angle averaged over all the unmasked pixels.

\section{\label{sec:axion} ALP signal}

We model the DM halo in a dwarf galaxy as a spherical system.
The possibility of a non-spherical halo in Leo T and Sculptor has been investigated, e.g., in \cite{Hayashi:2016kcy}. They found negligible impact in the derived D-factor uncertainties. We are not aware of non-spherical analysis for Eridanus 2, Grus 1, and Hydra II.
For simplicity, and given that we generically expect lower triaxiality for lower-mass DM halos~\cite{Schneider:2012}, we assume spherical symmetry for all the targets.

The flux density at wavelength $\lambda$ produced by decays of ALPs from a given direction identified by $\theta$ can be computed as:
\be
S_\lambda (\theta)=\frac{\Gamma_a}{4\pi}\,\frac{1}{\sqrt{2\pi}\sigma_\lambda} \exp{\left[-\frac{(\lambda-\lambda_{obs})^2}{2\sigma_\lambda^2}\right]}\int d\Omega\, d\ell \rho_a[r(\theta,\Omega,\ell)]\,B(\Omega)\;.
\label{eq:flux}
\ee

The decay rate $\Gamma_a$ depends on the ALP mass $m_a$ and the effective ALP-two-photon coupling $g_{a\gamma}$. In natural units, it reads
$\Gamma_a=g_{a\gamma}^2\,m_a^3/(64\pi)$.

The wavelength of emission can be computed as $\lambda_{em}=c/\nu_{em}$ with $\nu_{em}=m_a/(4\pi)$. In Eq.~\ref{eq:flux}, we neglect the velocity dispersion of ALPs in the dwarf halo, since it is typically $\lesssim 10^{-4}~c$, namely smaller than the spectral resolution of MUSE, $\sigma_\lambda/\lambda$ which varies between $1.2\times 10^{-4}$ and $2.7\times 10^{-4}$. However, we cannot neglect the heliocentric radial velocity of dwarfs, except for Leo T, and we correct for the Doppler shift when deriving the results below. The observed wavelength $\lambda_{obs}$ is then given by $\lambda_{obs} = \lambda_{em} (1 + v_{radial})$.

We assume a Gaussian behavior for both the energy and the angular responses of the detector, with FWHM as a function of the wavelength taken from Ref.~\cite{2017A&A608A1B} and normalized to the value at 7000 \AA\ mentioned in Sec.~\ref{sec:data}. The angular beam is denoted by $B(\Omega)$.

Under the assumption of spherical symmetry, the DM density $\rho_a(r)$ is a function only of the radial distance $r$ from the center of the dwarf, which can be expressed in terms of the coordinate along the line of sight $\ell$ and the angle of observation.
Our reference scenario for the description of the DM density profile is given by a ``cuspy'' distribution given by the NFW functional form~\cite{Navarro:1995iw}:
\be
\rho_{\rm{NFW}}(r)=\frac{\rho_s}{\left(\frac{r}{r_s}\right)\left( 1 + \frac{r}{r_s} \right)^2}\;,
\label{eq:rho}
\ee
where $\rho_s$ and $r_s$ are respectively the scale density and radius.
We also consider a second parameterization, dubbed ``coreNFWtides'', which modifies the cusp by allowing for a central core (e.g., due to star-formation feedback) and includes a decrease in density beyond a tidal radius~\cite{Read:2018pft}. This parametrization is not completely independent from the NFW one, since it builds on it, but it has the advantage to allow an easy assessment of the effect a core and a tidal radius have on our bounds. The coreNFWtides profile is described in more detail in the Appendix~\ref{sec:scu}.

We constrain the DM profiles of the dwarf galaxies through a Jeans analysis of the velocity dispersion of their stars. For Eridanus 2, Grus 1, Hydra II, and Leo T, we used the likelihood derived in \citep{Bas:2021} (GravSphere method). In the case of Sculptor, we derived the likelihood ourselves, using the same approach as in \citep{Bas:2021}, but with data from~\cite{2009AJ....137.3100W,Coleman_2005}.
The results for the Sculptor case are reported in Appendix~\ref{sec:scu}.
In the rest of our analysis, $\rho_s$ and $r_s$ are treated as free parameters of the model. For the coreNFWtides profile, the additional parameters describing the DM distribution, see Appendix~\ref{sec:scu}, are fixed to their global best-fit values from the just mentioned analyses of dispersion velocities, in order to reduce the number of free parameters.   We expect that a full scan of all the parameters of the coreNFWtides distribution would lead to a mild weakening of our bounds, as discussed in Appendix~\ref{sec:cav}.

The other parameters that enter Eq.~\ref{eq:flux} and that will be sampled in our scans are $g_{a\gamma}$ and $m_a$. In total, there are four free parameters describing the expected flux from ALPs.

\section{ Methods and Results}\label{sec:res}
In our statistical analysis, we consider two types of data. On one side, we have the dispersion velocities of the stellar component in the dwarf galaxy, which allow us to infer the DM spatial distribution via Jeans analysis. From the likelihood defined in~\citep{Bas:2021}, we derive a profile likelihood $\mathcal{L}_{Jeans}^j$ depending only on $\rho_s$ and $r_s$, where all the other ``nuisance'' parameters are profiled out (and, in the case of the coreNFWtides profile, the additional parameters are set to their global best-fit values).
The index $j$ stands for the dwarf considered.

The second type of dataset we consider is the diffuse emission probed by MUSE observations in the direction of the dwarf galaxies.
As done in \citep{Regis:2021}, we compare the expected ALP signal with the observed data in each dwarf by means of a Gaussian likelihood (omitting the index $j$ for simplicity):
\be 
\mathcal{L}_{diff}=e^{-\chi^2/2} \;\;\; {\rm with} \;\;\; \chi^2=\frac{1}{N_{pix}^{FWHM}}\sum_{i=1}^{N_{pix}} \left(\frac{S_{th}^i-S_{obs}^i}{\sigma_{rms}^i}\right)^2\;,
\label{eq:like}
\ee
where $S_{th}^i$ is the theoretical estimate for the flux density in the pixel $i$, $S_{obs}^i$ is the observed flux density and $\sigma_{rms}^i$ is the r.m.s. error, both described in \Sec{sec:data}. 
The theoretical estimate is given by Eq.~\ref{eq:flux} along with an additional spatially flat term $S_{\lambda,flat}$ that we incorporate in the fit to each individual map at every wavelength to account for incomplete sky subtraction. We consider this flat term a nuisance parameter.
$N_{pix}$ is the total number of pixels in the area under investigation, 
which we chose to be a circle of $60''$ of radius. The number of pixels within the MUSE angular beam is $N_{pix}^{FWHM}$, which has a size given by the aforementioned FWHM.

We define a likelihood $\mathcal{L}_{diff}$ which depends only on ALP parameters by profiling out $S_{\lambda,flat}^j$ from the likelihood in Eq.~\ref{eq:like}. Then, assuming the two types of datasets to be independent, we can define, at any given mass $m_a$, a global likelihood for each dwarf $j$:
\be 
\mathcal{L}^j(g_{a\gamma},\rho_s^j,r_s^j)=\mathcal{L}_{diff}^j(g_{a\gamma},\rho_s^j,r_s^j)\times \mathcal{L}^j_{Jeans}(\rho_s^j,r_s^j) \;,
\label{eq:likeall}
\ee
and a combined likelihood considering all five targets simultaneously:
\be
 \mathcal{L}^{all}(g_{a\gamma},\vec \rho_s,\vec r_s)=\prod_{j=1}^5 \mathcal{L}^j(g_{a\gamma},\rho_s^j,r_s^j)
\label{eq:likecomb}
\ee
To compute $\mathcal{L}_{diff}^j$,  we scan a three-dimensional logarithmically spaced grid of values of the following parameters: $k = g_{a\gamma}\sqrt{\rho_s}$, $r_s$ and $S_{\lambda, flat}$. $k$ is sampled in the range [$10^{-15}$, $10^{-10}$]~GeV$^{-1}(10^8~M_\odot~\mathrm{kpc}^{-3})^{1/2}$. The values of $g_{a\gamma}$ and $\rho_s$ are then separated by sampling $g_{a\gamma}$ in the range [$10^{-13.6}$, $10^{-10.5}$]~GeV$^{-1},$ while the ranges of variation of $\rho_s$ and $r_s$ are different for every galaxy and are determined from the 95\% C.L. contour of $\mathcal{L}^j_{Jeans}$. 
Lastly, for each channel, we first compute the best-fit value of $S_{\lambda, flat}$ for a vanishing $g_{a\gamma}$, and then we scan $S_{\lambda, flat}$ in a range from 1/10 to 10 times this best-fit value.

We assume that $\lambda_c(g_{a\gamma})=-2\ln[\mathcal{L}(g_{a\gamma},\vec \rho_s^{\,lbf},\vec r_s^{\,lbf})/\mathcal{L}(g_{a\gamma}^{b.f.},\vec \rho_s^{\,gbf},\vec r_s^{\,gbf})]$ follows a $\chi^2$-distribution with one d.o.f. and with one-sided probability given by $P=\int^{\infty}_{\sqrt{\lambda_c}}d\chi\,e^{-\chi^2/2}/\sqrt{2\,\pi}$, where $g_{a\gamma}^{b.f.}$ denotes the best-fit value for the coupling at a specific ALP mass. The superscript $gbf$ indicates the global best-fit, i.e. for $g_{a\gamma}=g_{a\gamma}^{b.f.}$, whilst $lbf$ denotes the best-fit of $\vec \rho_s$ and $\vec r_s$ for that given $g_{a\gamma}$. For the analysis on a single dwarf $j$, one has just to replace $(\vec \rho_s,\vec r_s)$ with $(\rho_s^j,r_s^j)$ in the expression of the estimator $\lambda_c$.
The 95\% C.L. upper limit on $g_{a\gamma}$ at mass $m_a$ is obtained from $\lambda_c=2.71$.

\begin{figure}[t!]
\centering
   \includegraphics[width=\textwidth]{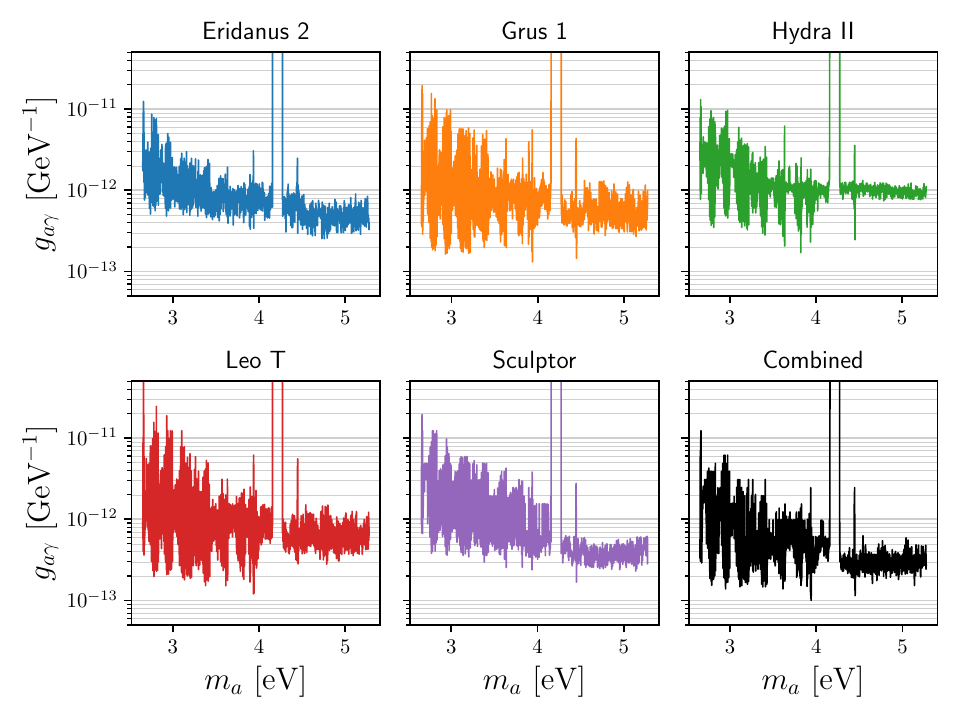}
    \caption{95\% C.L. upper limits on the effective ALP-two-photon coupling $g_{a\gamma}$ as a function of the ALP mass, derived in this work for the five dwarf spheroidal galaxies under investigation, and in the combined case (last panel). The limits are computed assuming an NFW spatial profile for the ALP DM.
}
\label{fig:bounds}
 \end{figure}

Results are shown in Figs.~\ref{fig:bounds} (NFW profile) and \ref{fig:bounds2} (coreNFWtides). We see that the different targets provide similar bounds, which also further motivates us to perform the combined analysis. The coupling $g_{a\gamma}$ is constrained at a level around $10^{-12}~\rm{GeV}^{-1}$ with significant fluctuations between adjacent masses, due to the noise from the process of subtracting the foreground emission lines.
Such rapid variation is more pronounced at lower masses/longer wavelengths reflecting the presence of strong OH emission lines from the night sky in this wavelength range.
The bounds improve slightly from low to high masses, which is due to the scaling of the decay rate with $m_a^3$, mitigated by an opposite energy dependence of the observational capabilities (angular and energy resolutions, foreground).

\begin{figure}[t!]
\centering
   \includegraphics[width=0.7\textwidth]{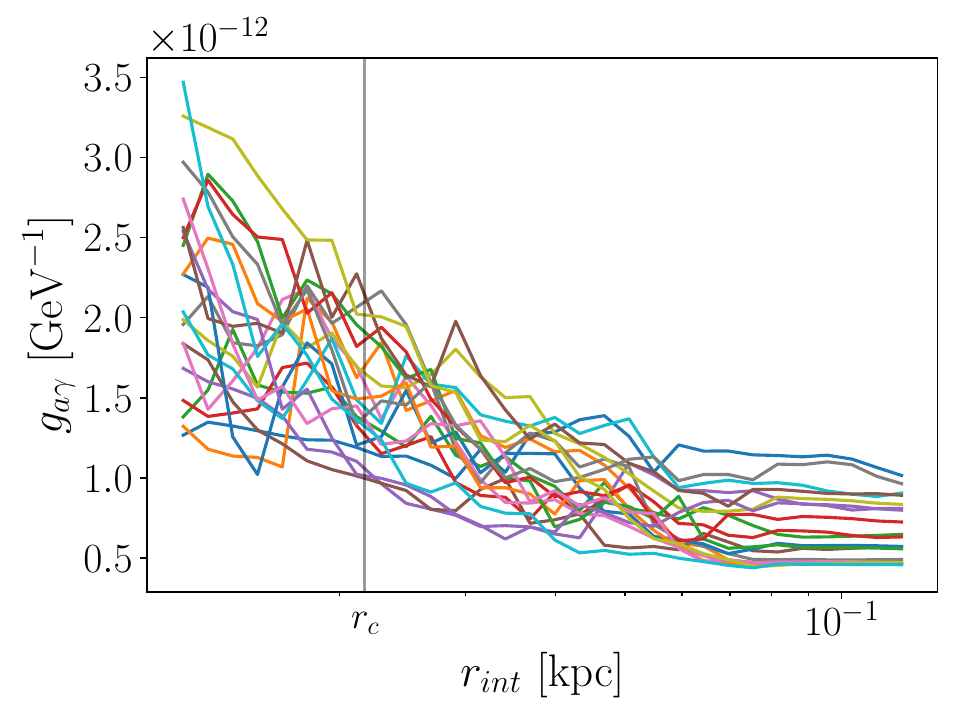}
    \caption{Bounds for the Leo T galaxy as a function of the integration radius for 20 channels with wavelengths equally spaced in the range [5324.57, 5562.07]~\AA,~ and using a coreNWFtides profile. The vertical line marks the best-fit value of the core radius.
}
\label{fig:rint_vary}
 \end{figure}
By comparing Figs.~\ref{fig:bounds} and \ref{fig:bounds2}, we see that the reduction of constraining power for the coreNFWtides profile is very limited. This is because we are probing a relatively large portion of the targets, implying that the majority of the pixels entering the statistical analysis are not from the central region, where the two profiles differ, but from distances where the two profiles basically coincide. To further illustrate this point, we show in Fig.~\ref{fig:rint_vary} how the bounds change as a function of the integration radius $r_{int}$, corresponding to the radius of the circular area used to compute the bound (we remind that the default value in our analysis corresponds to an angular radius of $60''$).
Each of the 20 lines corresponds to one channel, while the vertical line marks the best-fit value of the core radius $r_c$. As a general trend, the bound becomes more stringent once regions larger than the core radius are considered.

\begin{figure}[t!]
\centering
   \includegraphics[width=\textwidth]{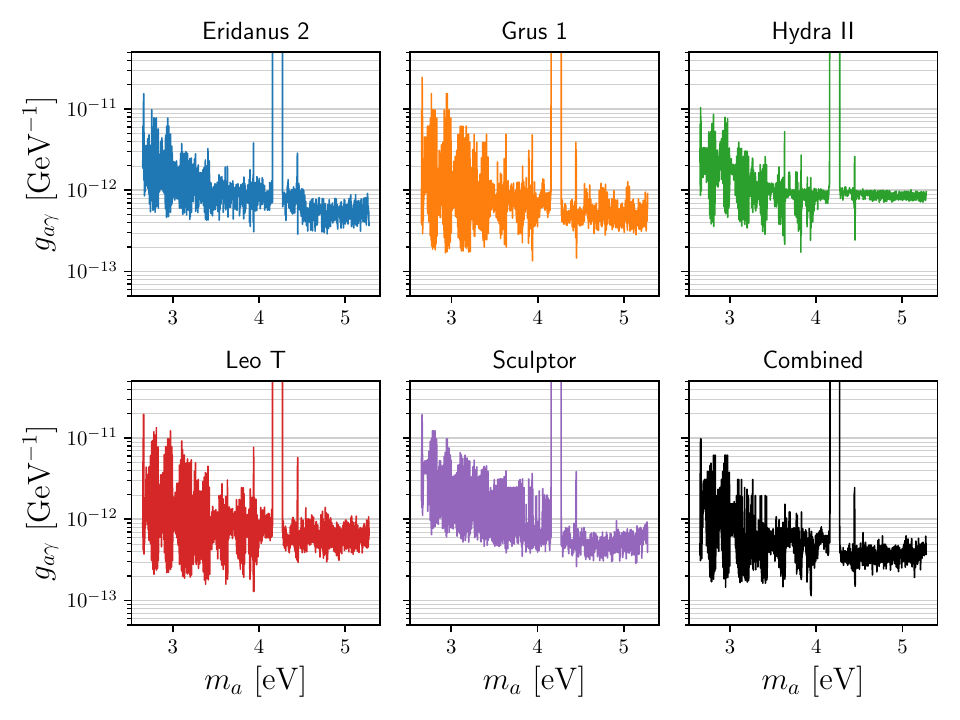}
    \caption{Same as Fig.~\ref{fig:bounds} but for the coreNFWtides DM profile. 
}
\label{fig:bounds2}
 \end{figure}

The robustness of our results against different masking and error estimates is tested in the same way as discussed in \citep{Regis:2021}. We find negligible differences in the derived bounds from the alternative analyses, with results very similar to what shown in Fig. 3 of \citep{Regis:2021}.

\begin{figure}[t!]
    \centering
    \includegraphics[width=0.9\textwidth]{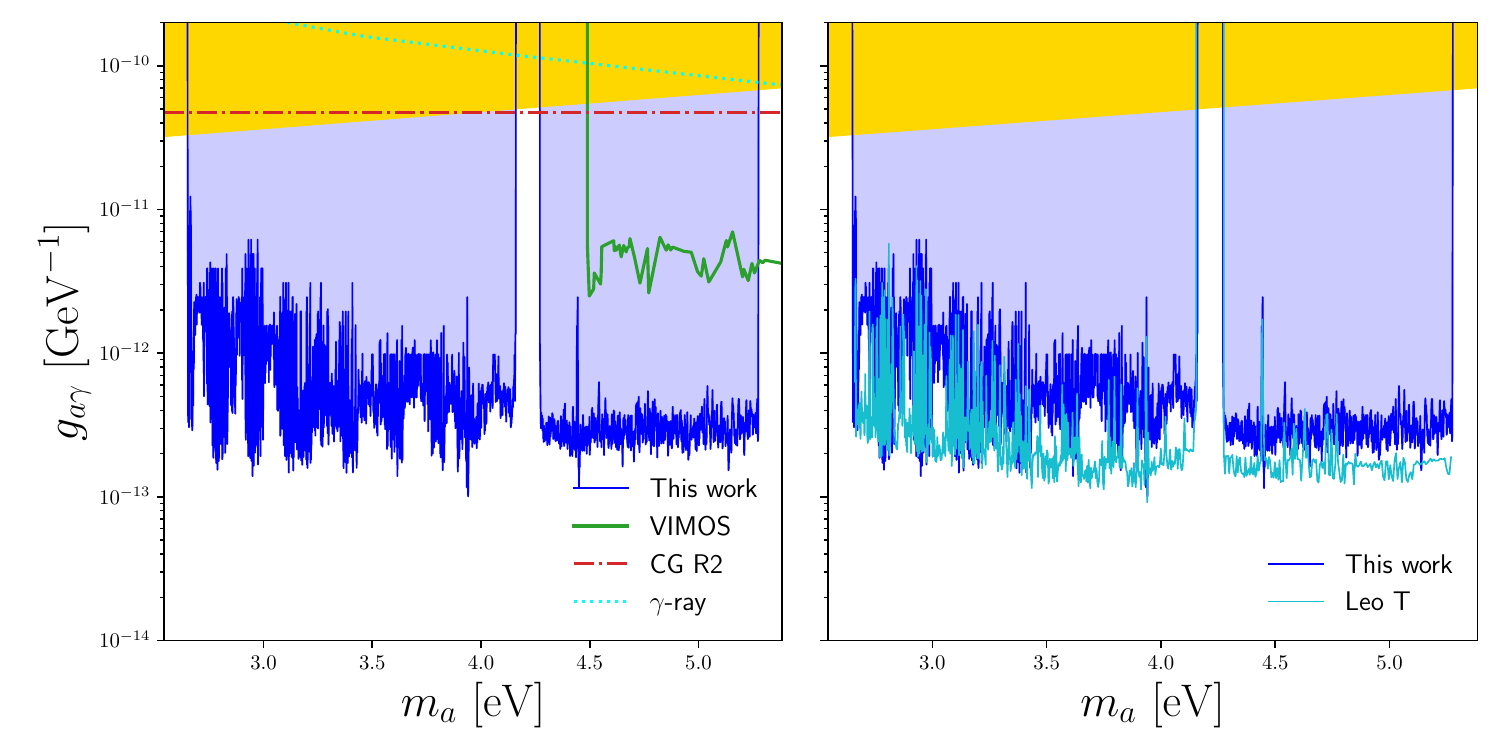}
    \caption{{\bf Left:} bounds derived in this work assuming an NFW profile exclude $g_{a\gamma}$ in the blue shaded region. The yellow region marks the QCD axion band~\cite{DiLuzio:2016sbl}. We show also bounds from star evolution in globular clusters from~\cite{Dolan:2022kul} (red dash-dotted line), from observation of galaxy clusters~\cite{Grin:2006aw} (green solid line), and the 99\% C.L. constraint from $\gamma$-ray attenuation~\cite{Bernal:2022xyi} (cyan dotted line).
    {\bf Right:} comparison with the bounds derived by the MUSE observation of Leo T in~\cite{Regis:2021}.}
    \label{fig:bounds_fin}
\end{figure}

In the left panel of Fig.~\ref{fig:bounds_fin}, we summarize our findings for the NFW profile and include the bound derived in Ref.~\cite{Grin:2006aw} from the observation of clusters, in Ref.~\cite{Dolan:2022kul} from the ratio of horizontal branch (HB) to Asymptotic Giant Branch stars in globular clusters and, for reference, the preferred region for the QCD axion~\cite{DiLuzio:2016sbl}. 
In the wavelength/mass range covered by our analysis, we can confidently exclude the QCD axion, which is also in tension with other astrophysical and laboratory probes associated with couplings different from $g_{a\gamma}$, see e.g.~\cite{Tanabashi:2018oca}, and the possible interpretation of near-infrared
background anisotropies in terms of ALP dark matter~\cite{Gong:2015hke}.

In the right panel of Fig.~\ref{fig:bounds_fin}, we compare the results of our combined analysis to Ref.~\cite{Regis:2021}, obtained from the MUSE data of Leo T.
The current analysis is typically more conservative, even though bounds are at a comparable level, and this is mainly due to the treatment of the DM profile. Indeed, in Ref.~\cite{Regis:2021}, the profile was derived by extrapolating results from a Jeans analysis at larger radii~\cite{Bonnivard:2015xpq}, while here it is derived directly from data. We found that, in the case of Leo T, the extrapolation slightly overshoots the real DM profile. On top of that, the uncertainty associated with the profile determination is now taken into account in a more rigorous statistical way, as described above.

For what concerns possible evidence of an ALP signal, we define, again at any given mass, $\lambda_d=2\ln[\mathcal{L}(g_{a\gamma}=0,\vec \rho_s^{\,lbf},\vec r_s^{\,lbf})/\mathcal{L}(g_{a\gamma}^{b.f.},\vec \rho_s^{\,gbf},\vec r_s^{\,gbf})]$.
The ALP discovery would occur if $\sqrt{\lambda_d}>5$.\\

Due to the imperfect subtraction of emission lines from the night sky, many channels present large values of $\sqrt{\lambda_d}$. In order to identify possible emission peaks due to the presence of an ALP, we need to remove this spurious evidence from our data, i.e. determine which channels are ``unreliable".\footnote{Note that, in those channels, the bounds described above are weakened and not strengthened by the presence of such residual emission.} As a first step, we look at the sky spectrum of Leo T, obtained from a data cube without sky subtraction, by summing the flux over an area with no bright sources. 
The sky spectrum presents large peaks above a non-zero frequency-dependent baseline level.
We search for peaks with heights above a threshold that we choose to be five times the standard deviation of fluctuations about the baseline in a region without large emission lines. With this criterion, we identify 225 peaks, all of which correspond to known atmospheric emission lines, except two, which we will not consider as sources of fake evidence in the following. Next, in order to determine how many channels are affected by a bright emission line, we construct an estimate of the reliability of the errors used in our analysis.

For this purpose, we use the individual Leo T exposures.  We reduce the data for 11 individual exposures, following the data reduction process described in Sec.~\ref{sec:data}. By assuming that the data are perfectly aligned across exposures, we have:\\ 
${f_i(\lambda) = f_{true}(\lambda) + N(0, \sigma_i(\lambda)^2)}$,
where $f_i(\lambda)$ is the flux measured at a given wavelength $\lambda$,  $f_{true}(\lambda)$ is the true flux at a given wavelength $\lambda$, and $N(0, \sigma_i(\lambda)^2)$ represents the normally distributed random noise at each wavelength, with variance $\sigma_i(\lambda)^2$. We use the spectra of 220 stars in the Leo T field of view, and, for each star, we compute $f_{true}(\lambda)$ as the average of the 11 observations (therefore $i$ runs over 220 stars and 11 exposures).
It is important to note that this modeling approach is only suitable for sources, as the sky background may vary. On the other hand, the fluxes from stars are relatively stable, and $f_{true}(\lambda)$ should be nearly constant.
If the uncertainty estimates $\sigma_i(\lambda)$ are reliable, then the difference between the observed flux and the true flux, normalized by $\sigma_i(\lambda)$, should be distributed approximately as a Gaussian with zero mean and standard deviation $\Sigma(\lambda) = 1$. 
Next, we superimpose the sky spectrum of Leo T with $\Sigma(\lambda)$. We observe that around the frequencies corresponding to atmospheric emission lines, $\Sigma$ deviates significantly from 1. To determine the characteristic range of wavelengths affected, we look at three well-known isolated oxygen emission lines
($\lambda_P = 5577.3,$ $6300.3,$ $6363.8~\rm{\AA}$). We find that $\Sigma(\lambda) - 1 > 0.05$  in a range $[\lambda_P-\delta\lambda,\lambda_P+\delta\lambda]$, with $\delta\lambda = 5.7 \sigma_\lambda$ and $\lambda_P$ being the wavelength of the peak and $\sigma_\lambda$ being the spectral resolution as in Eq.~\eqref{eq:flux}.

We thus conduct the evidence search excluding channels that fall into the $\pm \delta\lambda$ region around all background emission peaks. 
With this procedure, approximately 2000 channels out of 3719 are excluded.
Finally, we correct for the Look Elsewhere Effect (LEE) by dividing the p-value by a factor $N_{trials}$ equal to the total number of used channels divided by the number of channels falling within the spectral resolution. We compute the p-value from the test statistics $\lambda_d$ defined above, combining all targets.

We find no evidence for ALP DM in our data, i.e., no case with $\sqrt{\tilde{\lambda}_d}>5$, where $\tilde{\lambda}_d$ has the same meaning of $\lambda_d$ but corrected for the LEE.


\section{\label{sec:conc} Conclusions}
Most ALP models predict a coupling between photons and ALPs. 
This implies that we expect a monochromatic photon flux generated by ALP decays inside astrophysical structures. 
Nearby dwarf spheroidal galaxies are ideal targets for this search since they are DM-dominated and are relatively close to us.
Assuming ALPs to constitute all the DM in galaxy halos, we analyzed MUSE spectroscopic observations of five dwarf spheroidal galaxies to search for ALP radiative decays in the mass range 2.7-5.3~eV. 
The excellent spectral resolution and sensitivity of the spectroscopic observations obtained with the MUSE instrument at the VLT allowed us to probe quite faint and diffuse monochromatic line emissions.

We tested the possible presence of an ALP DM signal, concluding that none of the channels selected for this analysis, i.e., not affected by large background contamination, is exhibiting a detection. 
We derived robust bounds on the effective ALP-two-photon coupling consistent among the five galaxies considered. We took into account the uncertainties associated with the DM spatial distribution in each dwarf galaxy by including a profile likelihood depending on the DM profile parameters derived from the Jeans analysis of Ref.~\cite{Bas:2021}. We considered two possible profiles: an NFW and a cored profile, yielding comparable bounds. 
A different description of the DM profile from what considered here might lead to slightly different bounds. However, we do not expect large modifications since the constraining power comes from dSph regions where the determination of the DM density from kinematic data appears to be robust and so weakly dependent on the parameterization.

The resulting bounds lie well below the QCD axion band, and are significantly more constraining than limits from other probes, in the relevant mass range.

\section*{Acknowledgements}

MT acknowledges support from the research grant ‘The Dark Universe: A Synergic Multimessenger Approach’ No. 2017X7X85K funded by MIUR.
MR, JR, MT and ET acknowledge support from the project ``Theoretical Astroparticle Physics (TAsP)'' funded by the INFN.

MR, JR and ET acknowledge support from `Departments of Excellence 2018-2022' grant awarded by the Italian Ministry of Education, University and Research (\textsc{miur}) L.\ 232/2016 and Research grant `From Darklight to Dark Matter: understanding the galaxy/matter connection to measure the Universe' No.\ 20179P3PKJ funded by \textsc{miur}.
JB and DV acknowledge support by Fundação para a Ciência e a Tecnologia (FCT) through the research grants UIDB/04434/2020 and UIDP/04434/2020 and  through grant PTDC/FIS-AST/4862/2020. JB acknowledges work contract 2020.03379.CEECIND and DV acknowledges support from the Fundação para a Ciência e a Tecnologia (FCT) through the Fellowship 2022.13277.BD.

Based on observations made with ESO Telescopes at the La Silla Paranal Observatory under programme IDs 0100.D-0807, 0101.D-0300, 0102.D-0372 and 0103.D-0705.


\bibliographystyle{JHEP_improved}
\bibliography{biblio}

\appendix
\section{Jeans analysis of Sculptor}\label{sec:scu}
\begin{figure}[t]
\centering
\includegraphics[width=.5\linewidth]{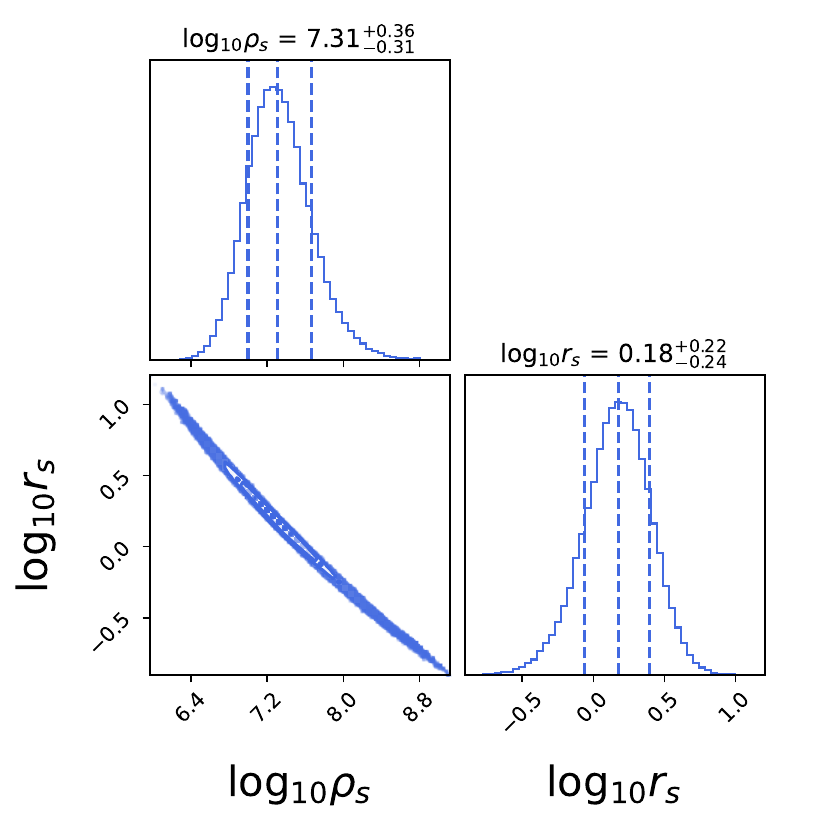}
\caption{Posterior distribution of the dark matter parameters entering the NFW profile in Eq.~\ref{eq:rho}. Units are $\log_{10}(\rho_s/(M_\odot/\rm{kcp}^3))$ and $\log_{10}(r_s/\rm{kpc})$.}
\label{fig:triangle_NFW}
\end{figure}

\begin{figure}[t]
\centering
\includegraphics[width=1\linewidth]{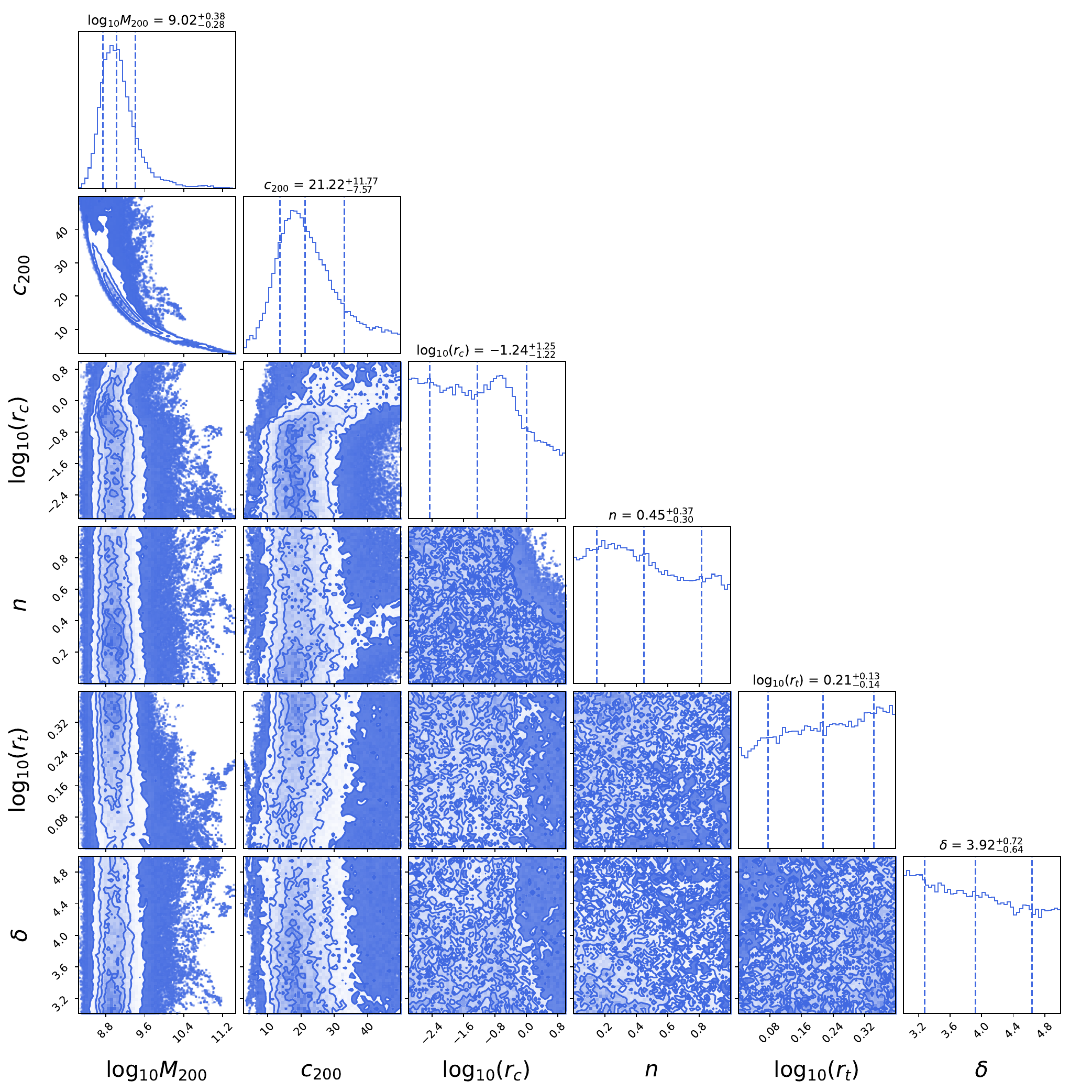}
\caption{The same as in Fig.~\ref{fig:triangle_NFW} but using a coreNFWtides profile. Units are $\log_{10}(M_{200}/(M_\odot))$, $c_{200}$, $\log_{10}(r_c/[\mathrm{kpc}])$, $n$, $\log_{10}(r_t/[\mathrm{kpc}])$, $\delta$.}
\label{fig:triangle_coreNFWtides}
\end{figure}

The data we have used in our Jeans analysis for Sculptor are the line-of-sight velocities reported in \cite{2009AJ....137.3100W} alongside the photometric measurements presented in \cite{Coleman_2005} for surface density data. We model the stellar dynamics assuming a spherically symmetric and non-collisional Jeans equation \cite{10.1093/mnras/190.4.873,10.1093/mnras/200.2.361}:
\be
\label{eq:Jeans}
\frac{1 }{\nu(r)}\frac{\partial}{\partial r} \left( \nu(r)\sigma_r^2\right) + \frac{2\beta(r)\sigma_r^2}{r} = \frac{GM(<r)}{r^2},
\ee
where $\nu(r)$ is the stellar density, $\sigma_r$ the radial velocity dispersion, $\beta(r)$ the velocity anisotropy and $M(<r)$ is the total enclosed mass within a radius $r$ from the center of the target. Furthermore, we model the radial stellar density profile as a sum of three Plummer spheres~\cite{10.1093/mnras/200.2.361}:
\be
\label{eq:stellar_den}
\nu(r) = \sum_{j=1}^{3}\frac{3 M_j}{4 \pi a_j^3} \left( 1 + \frac{r^2}{a_j^2} \right)^{-5/2},
\ee
$M_j$ and $a_j$ are parameters which can be constrained through the data.
The expression~\ref{eq:stellar_den} can be seen as a density expansion, analogous to a Gaussian decomposition. The velocity anisotropy $\beta(r)$ is parametrized as:
\be
\beta(r)=\beta_0 + (\beta_{\infty}-\beta_0)\,\frac{1}{1+\left( r_a/r \right)^\eta},
\ee
where $\beta_0$, $\beta_{\infty}$, $r_a$ and $\eta$ are also free parameters describing respectively the inner and outer orbital anisotropy, the radius and the sharpness of the transition.

The mass distribution is the sum of the DM component and the stellar contribution, which is modeled as in Eq.~\ref{eq:stellar_den} but with a free parameter fixing the overall normalization (in practice this corresponds to a free mass-to-light ratio).
As explained in \Sec{sec:axion}, we consider two options for the DM distribution, namely the NFW profile in Eq.~\ref{eq:rho}, and the coreNFWtides model of~\cite{Read:2018pft}, which modifies the NFW distribution allowing for the presence of a central core, and a reduced density beyond a tidal radius.
More specifically, the central core is implemented by modifying the NFW mass distribution as follows:
\begin{equation}
    M_{\rm{cNFW}}(<r)= M_{\rm{NFW}}(<r) f^n,
    \label{eq:core1}
\end{equation}
with $0\leq n \leq 1$ and 
\begin{equation}
    f=\tanh{\left( \frac{r}{r_c}\right)},
\end{equation}
being $r_c$ the size of the core.
The associated density profile is:
\begin{equation}
\rho_{\rm{cNFW}}(r)=f^n\,\rho_{\rm{NFW}}(r)+\frac{nf^{n-1}\left(1-f^2\right)}{4\pi r^2r_c} M_{\rm{NFW}}(<r).\label{coreNFWtides}
\end{equation}
Furthermore, since the galaxies that we are describing experience strong gravitational interactions with the host galaxy, tidal stripping is expected in their external regions.
Such an effect can be modeled by further modifying the enclosed mass beyond a tidal radius $r_t:$
\begin{equation}
    M_{\rm{cNFWt}}(<r) = 
    \begin{cases}
    M_{\rm{cNFW}}(<r) & \text{if $r < r_t$},\\
    M_{\rm{cNFW}}(r_t) + 4\pi \rho_{\rm{cNFW}}(r_t)\frac{r_t^3}{3-\delta}\left[ \left(\frac{r}{r_t}\right)^{3-\delta} -1 \right], & \text{if $ r > r_t$},
  \end{cases}
\end{equation}
which in terms of the density profile reads
\begin{equation}
    \rho_{\rm{cNFWt}}(r) = 
    \begin{cases}
        \rho_{\rm{cNFW}}(r)& \text{if $r<r_t$},\\
        \rho_{\rm{cNFW}}(r_t)\left( \frac{r}{r_t} \right)^{-\delta}, & \text{if $r>r_t$},
    \end{cases}
    \label{eq:core_tides}
\end{equation}
and the external slope $\delta$ is taken to be $\delta\geq3.$

Given all these ingredients, we can use the radial velocity dispersion obtained by solving equation \ref{eq:Jeans} to compute the line-of-sight velocity dispersion:
\be
\sigma_{\rm{L.O.S}}^2(R) = \frac{2}{\Sigma_* (R)}\int_R^\infty \left( 1 - \beta \frac{R^2}{r^2} \right) \frac{\nu(r)\sigma_r^2 r}{\sqrt{r^2 - R^2}}dr,  
\ee
being $\Sigma_*(R)$ the projected stellar surface density, which can be expressed as:
\be
\Sigma_*(R) = \sum_{j=1}^{3} \frac{M_j}{\pi a_j^2}\left(1 + \frac{R^2}{a_j^2} \right)^{-2}.
\ee

Finally, we have compared the model to the data by considering  the Gaussian likelihood $-2 \ln{\mathcal{L}}= \chi^2_{\rm{L.O.S}}+\chi_{\Sigma_*}^2+\chi^2_{\rm{VSP}_1}+\chi^2_{\rm{VSP}_2}$.
The first two terms correspond to the chi-squared for the line-of-sight velocity dispersion and surface density data, respectively.
The last two contributions are constructed from the fourth moments of the velocities distribution, known as the virial shape parameters $\rm{VSP1}_1$ and $\rm{VSP}_2,$ which can be computed using equations [20-23] from Ref.~\cite{2017MNRAS.471.4541R}.
See Refs.~~\cite{2017MNRAS.471.4541R,Reynoso-Cordova:2022ojo} for more details.
The model is based on a total of 13 (17) free parameters: 2 (6) for NFW (coreNFWtides), 4 for the velocity anisotropy, and 7 for the stellar component.
We have explored this parameter space through a Markov Chain Monte Carlo (MCMC) simulation. 
The numerical analysis is performed through the {\sc emcee} \citep{2013PASP..125..306F} sampler which is implemented by using  public python code {\sc pyGravSphere} \citep{2020MNRAS.498..144G}. 
Prior to the MCMC, the stellar model is fitted to the surface density data. Then, in the MCMC the corresponding parameters $a_j$ and $M_j$ are allowed to vary in a 50\% range around their previously determined surface density best-fit values. The dark matter profile \ref{eq:core_tides} was added to the already existent profiles in {\sc pyGravSphere} using the same priors as in \citep{Bas:2021}.

We present our results for the case of an NFW profile in Fig.~\ref{fig:triangle_NFW}, where we show the posterior probability distributions of $\log_{10}(\rho_s/(M_\odot/\rm{kpc}^3))$ and $\log_{10}(r_s/\rm{kpc}),$ alongside their 0.16, 0.5 and 0.84 percentile values. 

Analogous results are shown in Fig.~\ref{fig:triangle_coreNFWtides} for the coreNFWtides distribution.
In this case, in order to follow the numerical implementation of the public code  {\sc GravSphere}~\cite{2017MNRAS.471.4541R,Read:2018pft,2021MNRAS.505.5686C}, we parametrize the NFW profile through the concentration $c_{\Delta}$ and the mass $M_{\Delta},$ which are
related to $\rho_s$ and $r_s$ by the equations:
\begin{equation}
    c_\Delta=r_\Delta/r_s \;\;\;,\;\;\;
    \rho_s = \frac{\Delta \rho_c c_{\Delta}^3}{3\left( \ln{(1 + c_\Delta) - \frac{c_\Delta}{1+c_\Delta}} \right)},
    \label{eq:conc_param}
\end{equation}
where $\rho_c$ is the critical density of the Universe, and $\Delta=200$, which in turn defines $r_{200}$ as the radius at which the DM density in the halo is 200 times the critical density of the Universe
\begin{equation}
    r_{200} = \left( \frac{3}{800}  \frac{M_{200}}{\pi  \rho_c} \right)^{1/3}.
    \label{eq:r200}
\end{equation}
Taking into account the additional parameters described in Eqs.~\ref{eq:core1}-\ref{eq:core_tides}, the coreNFWtides profile is defined by six parameters: $\log_{10}(M_{200}/[M_\odot])$, $\log_{10}(r_{200}/[\rm{kpc}])$, $\log_{10}(r_c/[\rm{kpc}])$, $n$, $\log_{10}(r_t/[\rm{kpc}])$ and $\delta$.
We show their posterior distribution from our analysis in Fig.\ref{fig:triangle_coreNFWtides}. Given the large number of parameters, this case shows significant degeneracy among the model parameters. On the other hand, the two most relevant ones, $M_{200}$ and $c_{200}$ are suitably well constrained.

\section{Caveats of the coreNFWtides analysis}\label{sec:cav}
\begin{figure}[t]
\centering
\includegraphics[width=.7\linewidth]{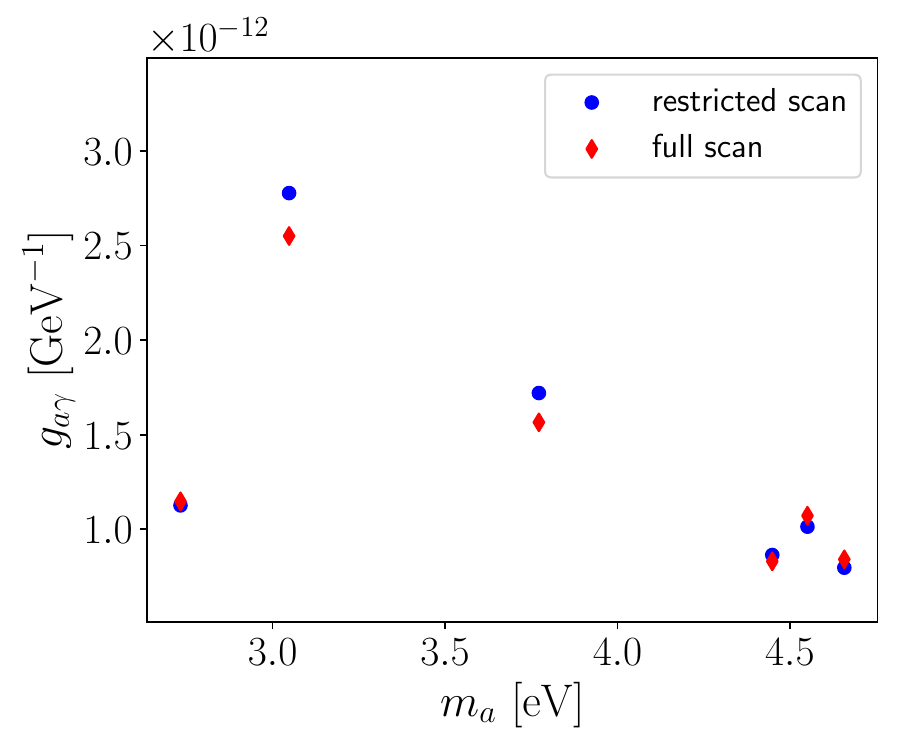}
\caption{Comparison between bounds obtained assuming a coreNFWtides profile for the Leo T galaxy performing a full scan of the profile parameter space (red diamonds) and a partial scan with four of the profile parameters fixed to their best-fit values as in Figure~\ref{fig:bounds2} (blue dots).}
\label{fig:fullscan}
\end{figure}

For the derivation of the bounds of Figure~\ref{fig:bounds2} under the assumption of the coreNFWtides profile Eq.~\eqref{coreNFWtides}, we have fixed $r_c$, $n$, $r_t$, and $\delta$ to their best-fit values from the Jeans analysis of Ref.~\cite{Bas:2021}. Although, in principle, all the parameters defining the profile distribution should be varied, the computational cost of scanning the eight-dimensional (six profile parameters, $g_{a\gamma}$ and $S_{flat}$) parameter space would render the analysis very expensive.
In particular, the bottleneck is the calculation of the likelihood from the MUSE data Eq.~\eqref{eq:like}, which involves a sum over a large number of pixels.

As discussed in Section~\ref{sec:res}, most of the constraining power comes from intermediate regions, located outside the core radius of the coreNFWtides profile as shown in Figure~\ref{fig:rint_vary}. In these regions the difference between the density profiles accounted for with our simplified procedure and with a full scan of all the parameters is limited.

To further check the validity of this argument, we perform a full parameter space scan for 6 channels of Leo T. The results are shown in Figure~\ref{fig:fullscan}. As expected, the change in the bounds compared to Figure~\ref{fig:bounds2} is minor.

\end{document}